\documentstyle[prl,aps,psfig,multicol]{revtex}

\def\lcmo{$\rm La_{0.67}Ca_{0.33}MnO_3$}
\def\lsmo{$\rm La_{0.67}Sr_{0.33}MnO_3$}

\begin{document}

\title{Charge carrier density collapse in ${\rm{}La_{0.67}Ca_{0.33}MnO_3}$ and
${\rm{}La_{0.67}Sr_{0.33}MnO_3}$ epitaxial thin films}

\author{W.~Westerburg, F.~Martin, and G.~Jakob}

\address{Institut f\"ur Physik, Johannes Gutenberg-Universit\"at Mainz,  D-55099 Mainz, Germany}

\author{P.J.M.~van Bentum and J.A.A.J.~Perenboom}

\address{High Field Magnet Laboratory, University of Nijmegen, 6525 ED Nijmegen, The Netherlands}

\date{22 July 1999}

\maketitle

\begin{abstract}
We measured the temperature dependence of the linear high field Hall resistivity
$\rho_H$ of {$\rm La_{0.67}Ca_{0.33}MnO_3$} ($T_C=232$\,K) and
{$\rm La_{0.67}Sr_{0.33}MnO_3$} ($T_C=345$\,K) thin films in the temperature range from 4\,K up to 360\,K in
magnetic fields up to 20\,T. At low temperatures we find a charge-carrier density of 1.3 and 1.4 holes per unit cell
for the Ca- and Sr-doped compound, respectively. In this temperature range electron-magnon 
scattering contributes to the longitudinal resistivity.
At the ferromagnetic transition
temperature $T_C$ a dramatic drop in the number of current carriers $n$ down to 0.6 holes per unit cell,
accompanied by an increase in unit cell volume, is observed.
Corrections of the Hall data due to a non saturated magnetic state will lead a more pronounced charge carrier density
collapse.
\end{abstract}


%
\begin{multicols}{2}
\section{ Introduction}
The colossal magnetoresistance (CMR) in ferromagnetic perovskite manganites has reattracted strong theoretical
and experimental interest. Experimental evidence exists that the origin of such a behaviour is the presence of
magnetic polarons. This concept of dynamic phase segregation is similar to the copper-oxide superconductors.
Small-angle neutron scattering measurements and magnetic susceptibility data on manganites reveal small
ferromagnetic clusters in a paramagnetic background
\cite{DeTeresa97}. Detailed  magnetotransport measurements in the paramagnetic phase well above the Curie temperature
confirm this picture \cite{Jakob98_2}. In the temperature and magnetic field range where the longitudinal transport
shows a negative temperature coefficient an electronlike, thermally activated Hall coefficient was found.
This is in agreement with the expectations from polaron hopping. However, for a given magnetic field
lowering of the temperature will lead to the formation of a polaron band and a 'metallic' transport, i.e. positive
temperature coefficient. In the vicinity of the metal-insulator transition the polaronic bands are stabilised
by external magnetic fields. Therefore we investigated the linear high field Hall resistivity in
high magnetic fields up to 20\,T.
The experimentally determined increase of the Hall coefficient
at the metal-insulator transition translates into a charge carrier density collapse (CCDC) in the band picture.
While such a CCDC in low magnetic field was proposed by Alexandrov and Bratkovsky \cite{Alexandrov99}
due to formation of immobile bipolarons, our high field results indicate the influence of the
structural phase transition at the Curie temperature on the band structure.
\section{ Experimental}
Thin films of La$_{0.67}$Sr$_{0.33}$MnO$_3$ (LSMO) were prepared by pulsed laser deposition
(KrF Laser, $\lambda=248$\,nm).
As substrates we used (100) SrTiO$_3$ and (100) LSAT [(LaAlO$_3$)$_{0.3}$-(Sr$_2$AlTaO$_6$)$_{0.7}$ untwinned].
The optimised
deposition conditions were a substrate temperature of 950$^{\circ}$C in an oxygen
partial pressure of 14\,Pa and annealing after deposition at 900$^{\circ}$C for 1 h in an oxygen partial pressure
of 600\,hPa. La$_{0.67}$Ca$_{0.33}$MnO$_3$ (LCMO) was deposited by magnetron sputtering on (100) MgO substrates.
Further details on preparation and characterisation are published elsewhere \cite{Jakob98_3}.
In X-ray diffraction in Bragg-Brentano geometry only film reflections corresponding to a ($l$00) orientation of
the cubic perovskite cell are visible for both compounds. The LSAT substrates ($a_0=3.87$\,\AA) have a low lattice
mismatch to the Sr-doped films
(3.89\,\AA). Rocking angle analysis shows epitaxial $a$-axis oriented growth with an angular spread smaller than
0.03$^{\circ}$. The in-plane orientation was studied by $\phi$-scans of (310) reflections. The cubic perovskite axes
of the films are parallel to that of the substrates with an angular spread smaller than 1\,$^{\circ}$.
The temperature dependence of the unit cell volume of the LCMO sample was determined measuring
in- and out of plane lattice constants using a helium flow cryostat with Beryllium windows
on a four circle x-ray diffractometer.\\
The samples were patterned photolithographically to a Hall bar structure. For measuring in the temperature regime
from 4\,K up to
room temperature we used a standard 12\,T magnet cryostat. An 8\,T superconducting coil and a 20\,T
Bitter type magnet
system, both with room temperature access, have been used for measurements above 270\,K. The procedure used for
measuring the Hall effect is described in detail elsewhere \cite{Jakob98}.
The magnetic moments of the films were measured with a SQUID magnetometer
in a small field of $B=20$ mT.
\section{ Results and Discussion}
In Fig.\ 1, the longitudinal resistivities of the Ca- and Sr-doped compounds $\rho_{xx}(T)$ are shown in zero field
(solid lines) and in 8\,T (dashed lines).
The Curie temperatures of both samples are indicated by arrows, 232\,K for LCMO and 345\,K for LSMO, respectively. For
LCMO the maximum in resistivity is close to $T_C$.
For very high and very low temperatures the curves are asymptotic, i.e.\ the magnetoresistance vanishes. The 
resistivity as function of temperature is up to $T/T_C=0.6$ given by 
\begin{equation}
    \label{Kubo}
    \rho=\rho_0+\rho_2T^2+\rho_{4.5}T^{4.5}.
\end{equation}
The parameters of Eq.\ \ref{Kubo}, obtained by fitting the experimental data, are listed in Table 1.
The quadratic contribution is not changed in presence of a high magnetic field. 
This was also observed by Snyder {\it et al.} \cite{Snyder96} while
Mandal {\it et al.} \cite{Mandal98} found both factors to be magnetic field dependent.
One possible origin of a quadratic temperature dependence of the resistivity is 
due to emission and absorption of magnons. But in this case a magnetic field
dependence would be expected. Furthermore in this processes an electron reverses its 
spin and changes its momentum. However, in the manganites spin flip 
processes play no role at low temperatures due to strong spin splitting of the states.
Therefore we attribute this $T^2$ dependent term 
to electron-electron scattering in a Fermi liquid.
The term proportional 
to $T^{4.5}$ results from electron-magnon scattering in the double exchange theory of 
Kubo and Ohata \cite{Kubo72}. In these scattering events the electron spin is conserved,
while momentum is exchanged between the electron and magnon system. 
Its contribution to the resistivity in Eq.\ \ref{Kubo} is
\begin{equation}
    \label{Kubo4_5}
    \rho_{4.5} = \frac{\epsilon_0\hbar}{e^2k_F} \frac{1}{S^2} (ak_F)^6 \left(\frac{m}{M}\right)^{4.5} 
                 \left(\frac{k_B}{E_F}\right)^{4.5}
\end{equation}
with the manganese spin $S$ and the lattice constant $a$. A small correction relevant 
only for effective mass ratio $M/m>1000$ is neglected in Eq.\ \ref{Kubo4_5}. The strong temperature dependence
of this scattering mechanism is partly determined by the number density of excited 
magnons $\propto T^{3/2}$. 
In magnetic field this term is suppressed reflecting the increased magnetic order.
Evaluating $\rho_{4.5}$ in the free electron approximation yields a value 3 orders of magnitudes
smaller than experimentally determined. However, due to the strong influences of effective mass
renormalizations in Eq.\ \ref{Kubo4_5} effective mass ratios of $M/m$ in the range 3-6 will 
give quantitative agreement.\\
For the Ca-doped compound the transport above $T_C$ is thermally activated \cite{Jakob98_2} and can be
described by small polaron hopping \cite{Emin69}.
The Sr-doped compound has a lower resistivity and a lower magnetoresistance (MR), because the absolute
MR decreases with increasing $T_C$ \cite{Khazeni96}. Here, the resistivity above $T_C$ is described by a crossover between
two types of polaron conduction \cite{Snyder96}. Scattering of polarons by phonons just above $T_C$ results in a
positive $d\rho/dT$, so that in the case of LSMO $T_C$ does not coincide with the maximum in resistivity. The Curie
temperature is at a lower value (see Fig.\ 1) where an anomaly in the temperature dependence of the resistivity is
seen \cite{Tokura94}.\\
The transverse resistivity in a ferromagnet as a function of a magnetic field $B$ is expressed by
\begin{equation}
    \label{RhoHall}
    \frac{{\rm d}\rho_{xy}}{{\rm d}B}=\frac{t}{I}\frac{{\rm d}U_H}{{\rm d}B}=R_H+R_A\mu_0\frac{{\rm d}M}{{\rm d}B} ,
\end{equation}
with the Hall voltage $U_H$, film thickness $t$, current $I$, magnetisation $M$, ordinary Hall coefficient $R_H$, and
anomalous Hall coefficient $R_A$ \cite{Karplus54}.
The Hall resistivity $\rho_{xy}(B)$ for LSMO for several constant temperatures as a function of magnetic field $B$
is shown in Fig.\ 2.
At low magnetic fields a steep decrease of the Hall voltage is seen, which is strongest at $T_C$ and becomes
less pronounced at low temperatures and above $T_C$.
This part is dominated by the increase in magnetisation with magnetic field. Therefore the electronlike anomalous Hall
effect, $R_A > R_H$, dominates the Hall voltage.
At higher fields the magnetisation
saturates and a linear positive slope due to the ordinary Hall effect is seen. This behaviour is very similar to the
Ca-doped compound, if one compares the reduced temperatures $T/T_C$ \cite{Jakob98_2}.
The initial slopes ${\rm d}\rho_{xy}/{\rm d}B (B\rightarrow{}0)$ are highest at the Curie-temperature for both compounds
in agreement with the Berry phase theory of the anomalous Hall effect \cite{Ye99}. The temperature
dependence of the electronlike anomalous Hall constant is thermally activated, similar to the longitudinal resistivity
and consistent with the theory of Friedman and Holstein \cite{Friedman63}.
This was also observed by Jaime {\it et al.}
\cite{Jaime97} and provides another strong evidence of small polarons in manganites. More details to the interpretation
of the anomalous Hall effect are published elsewhere \cite{Jakob98_2,Jakob98}.\\
In the following, we consider the high-field regime where the slopes ${\rm d}\rho_{xy}/{\rm d}B$
are positive and constant indicating hole conduction.
 For 4\,K, we obtain in a single-band model
1.4 holes per unit cell for LSMO. A smaller value $\approx$ 1 hole per unit cell was found by Asamitsu and Tokura
on single crystals \cite{Asamitsu98} while for a thin film a value of 2.1 was reported \cite{Snyder96}. For
Ca-doped thin films similar values were found \cite{Jakob98,Matl98}. This large charge carrier density in
manganites requires the concept of a partly compensated Fermi surface \cite{Pickett97}.\\ 
To investigate the
charge carrier concentration just above $T_C$, it is important to have as high magnetic field as possible in
order to saturate the magnetisation.  
Therefore, we performed for LCMO Hall effect measurements up to 20\,T. In this field a
positive linear slope  ${\rm d}\rho_{xy}/{\rm d}B$ can be observed up to $T/T_C=1.3$. The Hall resistivites
$\rho_{xy}(B)$ are shown in Fig.~3. For clarity several curves ($T=$ 275\,K, 300\,K, 305\,K, 310\,K and 315\,K) are
omitted. Just above $T_C$ at 285\,K it is yet possible to almost saturate the magnetisation of the sample 
and a broad field range with a linear positive slope remains to
evaluate $R_H$. The fit of the slope is seen in Fig.\ 3 by the dashed line. 
At 350\,K ($T/T_C=1.5$) the shift of the minimum in the Hall voltage to higher fields allows 
no longer a quantitative analysis of $R_H$.\\ 
Assuming full saturation of the magnetisation in high magnetic field the charge carrier concentration $n=1/eR_H$ as a
function of the reduced temperature $T/T_C$ for LCMO (circles) and LSMO (triangles) is plotted in Fig.~4. LCMO has
a constant carrier concentration at low temperatures, whereas for LSMO $n$ increases with temperature. But for
both doped manganites clearly a decrease of $n$ at the Curie temperature is seen. This temperature dependence of
$n$ seems to be a characteristic behaviour of the manganites in the vicinity of $T_C$, since this was also
observed by Wagner  {\it et al.} \cite{Wagner98} in thin films of Nd$_{0.5}$Sr$_{0.5}$MnO$_3$, indirectly by Ziese
{\it et al.} \cite{Ziese99} in thin films of La$_{0.67}$Ca$_{0.33}$MnO$_3$ and La$_{0.67}$Ba$_{0.33}$MnO$_3$ and
by Chun {\it et al.} \cite{Chun99} in single crystals of La$_{0.67}$(CaPb)$_{0.33}$MnO$_3$. The latter data are
also shown as open symbols in this figure for comparison. According to Fig.\ 4 above $T_C$ the number of charge
carriers seems to reincrease. The error bars indicate
the accuracy of the determination of the linear slopes ${\rm d}\rho_{xy}/{\rm d}B$, as shown for $T=285$ K in
Fig.\ 3. 
However, well above the Curie temperature the magnetisation of the sample cannot be saturated in  
experimentally accessible magnetic fields. From Equation \ref{RhoHall} it is obvious, that a linearly 
increasing magnetisation in this paramagnetic regime will not affect the linearity of the slope ${\rm d}\rho_{xy}/{\rm d}B$ but change its value.
Without quantitative knowledge of the anomalous Hall coefficient $R_A$ and the sample magnetisation $M(T,B)$ it is
not possible to separate this contribution. Nevertheless, since the sign of the anomalous Hall contribution is
electronlike, it is clear that the apparent charge carrier density shown in Fig.\ 4 has to be corrected to {\em
lower} values above the Curie temperature, thus {\em enhancing} the CCDC.\\
This CCDC indicates strong changes in the electronic distribution function close to the Fermi energy.
Since coincidence of structural, magnetic and electronic phase transitions has been reported
for La$_{1-x}$Ca$_{x}$MnO$_3$ with $x=0.25$ and 0.5 \cite{Radaelli95}, we investigated
the temperature dependence of lattice constants, magnetisation, longitudinal resistivity, and transversal resistivity
for the same sample. Fig.\ 5 shows a compilation of the results.
The charge carrier density is constant up to 0.7 $T_C$. In this temperature range
the resistivity follows Eq.\ \ref{Kubo} and the volume of the unit cell increases
slowly with temperature. The fact that the CCDC is accompanied by a strong increase in unit cell volume
and longitudinal resistivity
and a decay of the spontaneous magnetisation shows the coincidence of structural, magnetic, and
electronic phase transitions in La$_{0.67}$Ca$_{0.33}$MnO$_3$. \\

Since the transport in the manganites above $T_C$ is dominated by polaron hopping \cite{Jakob98_2,Jaime97}
we want to discuss the relation between our experimental observation of the CCDC and
the polaronic CCDC as proposed recently by Alexandrov and Bratkovsky \cite{Alexandrov99}.
They worked out the theory for a CCDC due to a phase transition of mobile polarons
to immobile bipolaronic pairs. At low temperatures the charge carriers form a
polaronic band. With increasing temperature the polaronic bandwidth decreases due to the increase
in electron phonon coupling. Depending on the polaron binding energy and the doping level a first or second
order phase transition to a bound bipolaronic state takes place. At still higher temperatures
thermal activation of the bipolarons leads to a reincrease of the number of mobile polarons.
Indeed their calculated temperature dependence of the number density of mobile polarons
in zero field is very similar to our data shown in Fig.\ 4.
However, in this model the density of mobile polarons around $T_C$ is a strong function of
magnetic field, which is responsible for the colossal magnetoresistivity.
In high fields the polaronic CCDC is strongly reduced, since the formation of bound bipolarons is suppressed
and accordingly the number of mobile polarons remains almost constant at $T_C$.
Therefore our observation of a CCDC in high fields cannot be identified with the
CCDC due to bipolaron formation, but is related to structural changes.
We cannot determine the charge carrier density in the low field regime due to strong anomalous Hall
contributions. In their presence the type of phase transition proposed by Alexandrov and Bratkovsky
cannot be verified by Hall effect measurements.\\

\section{ Summary}
We performed detailed Hall effect measurements in high magnetic fields in LCMO and LSMO thin films.
The charge carrier concentration was investigated as a function of temperature below and above the Curie temperature.
In the low temperature range, where the charge carrier density is constant, we identified electron-magnon 
scattering in the longitudinal resistivity.
At the ferromagnetic transition temperature a charge carrier density collapse was observed for both compounds. The
data indicate a simultaneous structural, magnetic and electronic phase transition
in doped manganite thin films. 
\acknowledgments
This work was supported by the Deutsche Forschungsgemeinschaft through project JA821/1-3 and the
European Union TMR-Access to Large Scale Facilities Plan.

\begin{figure}
    \centerline{\psfig{file=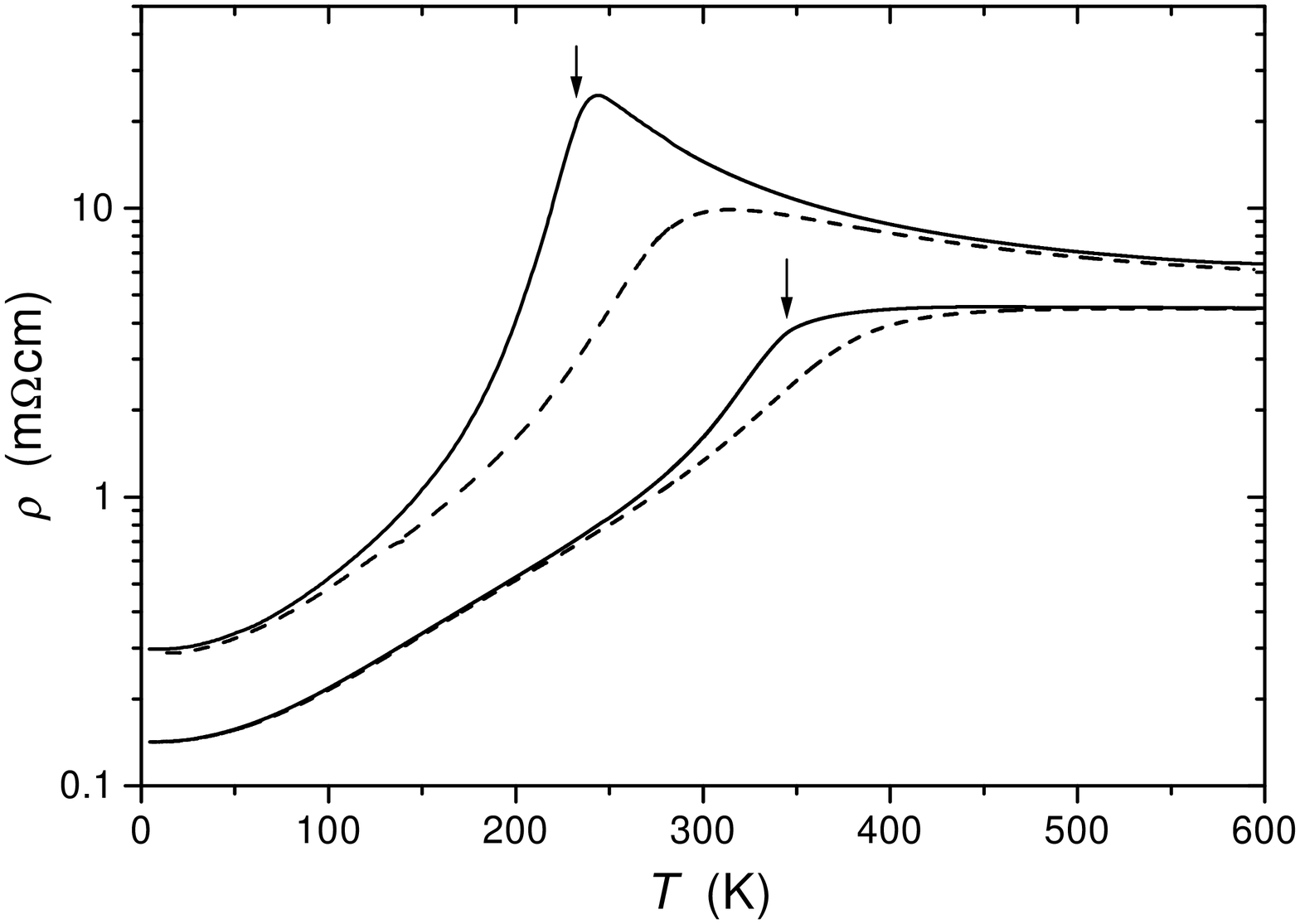,width=1.0\columnwidth}}
    \caption{Resistivities in zero field (solid lines) and in 8\,T magnetic field (dashed lines) as functions
    of temperature. The arrows at 232\,K (345\,K) indicate the Curie temperatures for La$_{0.67}$Ca$_{0.33}$MnO$_3$
    (La$_{0.67}$Sr$_{0.33}$MnO$_3$).}
    \label{fig1}
\end{figure}
\begin{figure}
    \centerline{\psfig{file=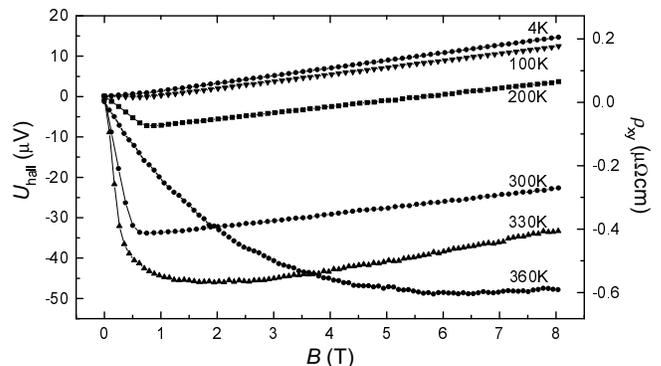,width=1.0\columnwidth}}
    \caption{The Hall resistivity $\rho_{xy}$ and the Hall voltage $U_H$ as a function of applied magnetic field
    at various temperatures for La$_{0.67}$Sr$_{0.33}$MnO$_3$.}
    \label{fig2}
\end{figure}
\newpage
\begin{figure}
    \centerline{\psfig{file=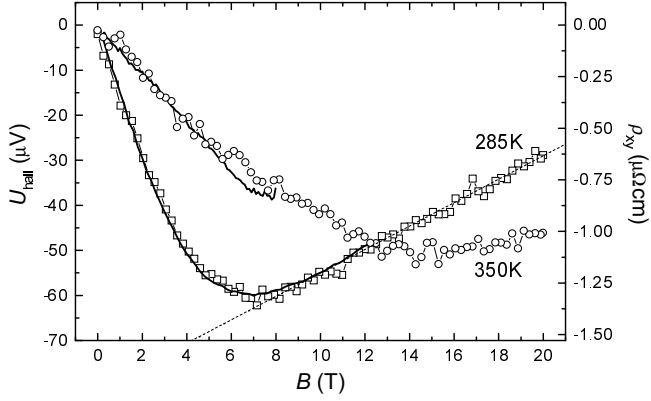,width=1.0\columnwidth}}
    \caption{Measured Hall resistivity $\rho_{xy}$ and Hall voltage $U_H$ as function of magnetic field
    at temperatures of 285\,K (squares) and 350\,K (circles) for La$_{0.67}$Ca$_{0.33}$MnO$_3$ in a
    20\,T Bitter magnet.
    Five measured curves were omitted for clarity. The lines are measurements in a 12\,T superconducting
    magnet and show the correspondence between different experimental setups. The dotted line is a fit
    of the linear slope.}
    \label{fig3}
\end{figure}
\begin{figure}
    \centerline{\psfig{file=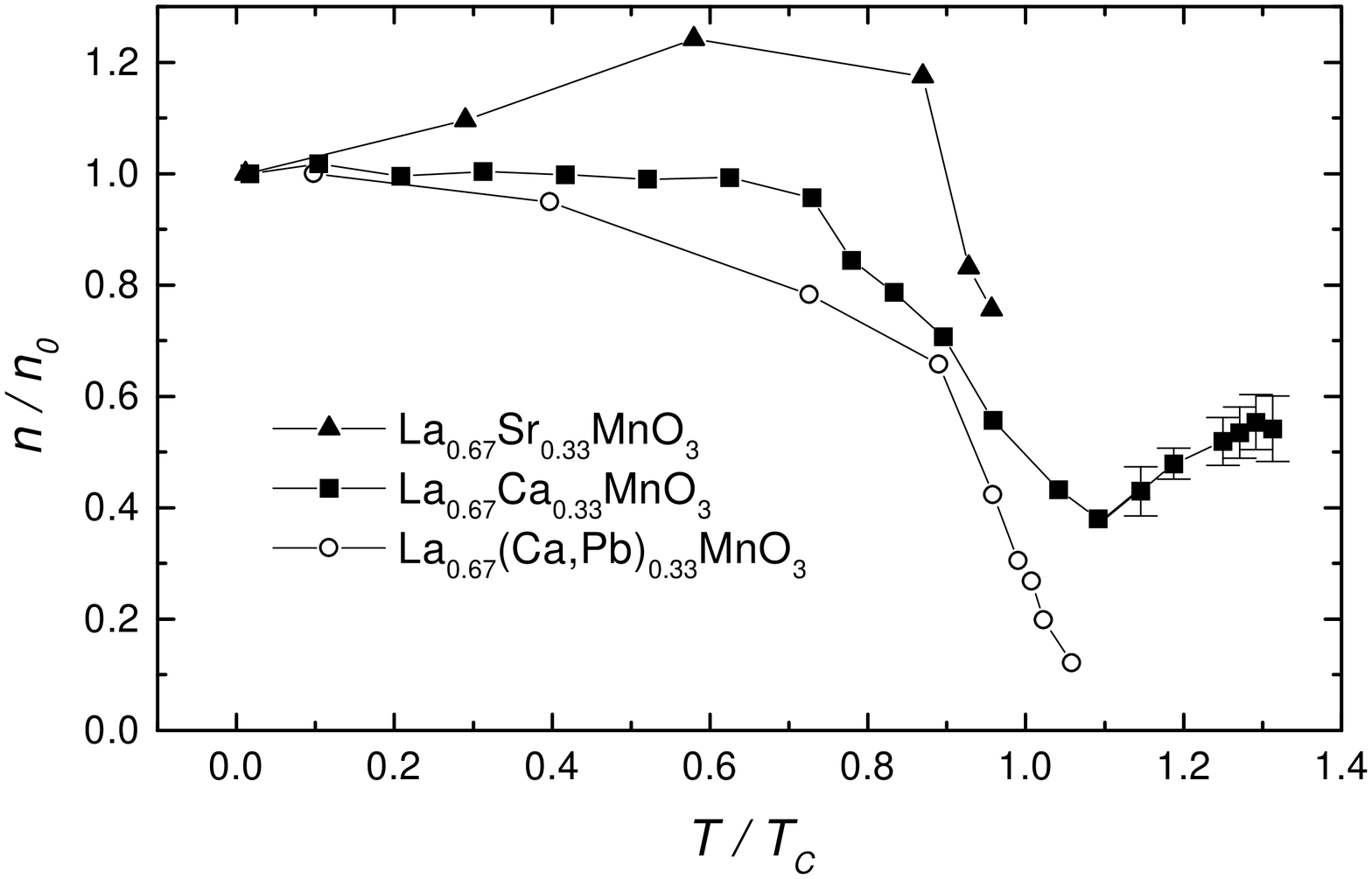,width=1.0\columnwidth}}
    \caption{Apparent normalised charge-carrier density
        $n \propto ({\rm d}\rho_{xy}/{\rm d}B)^{-1}$ of \lcmo\ (filled squares) and \lsmo\
    (filled triangles) as a function of the reduced temperature $T/T_C$.
    For the measurements in
        the 12\,T superconducting magnet errors are comparable to symbol size. In the 20\,T Bitter magnet
        electrical noise leads to enhanced error bars for the linear slopes $({\rm d}\rho_{xy}/{\rm d}B)^{-1}$
        as indicated.
        Additionally we show data from
        Ref.\ [21] 
        (open circles) on a single crystal of La$_{0.67}$(CaPb)$_{0.33}$MnO$_3$. }
    \label{fig4}
\end{figure}
\begin{figure}
    \centerline{\psfig{file=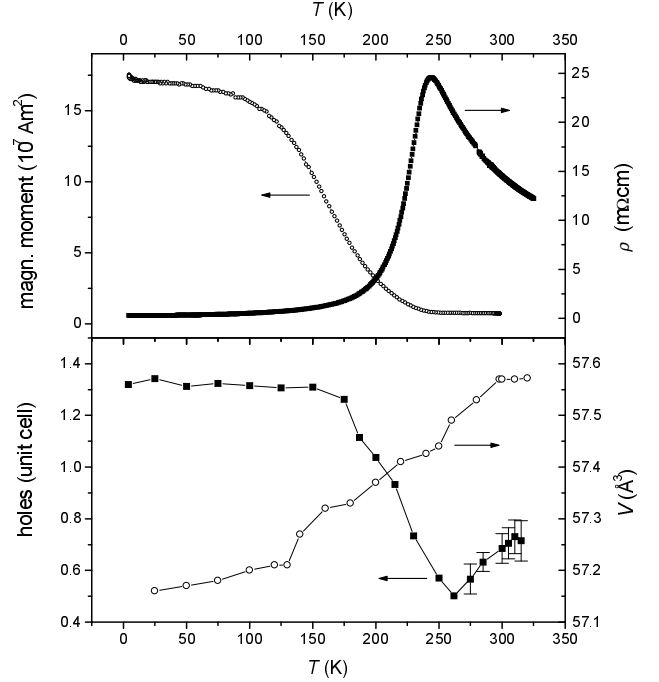,width=1.0\columnwidth}}
    \caption{Temperature dependence of magnetisation, resistivity, charge carrier density and unit cell volume
        for a thin film of La$_{0.67}$Ca$_{0.33}$MnO$_3$.}
    \label{fig5}
\end{figure}
\end{multicols}

\begin{table}[t]
\begin{tabular}{|c||c|c|c|c|c|c|}
Compound & $T_C$ & $n$(4 K)& $\rho_0$ & $\rho_2/\rho_0$ & $\rho_{4.5}(0\ \rm{T})/\rho_0$ & $\rho_{4.5}(8\ \rm{T})/\rho_0$\\
         &  (K)  & (holes/unit cell)   & (m$\Omega$cm) & ($10^{-6}\ \rm{K}^{-2}$)  & ($10^{-11}\ \rm{K}^{-4.5}$) & ($10^{-11}\ \rm{K}^{-4.5}$)\\                  \hline
LCMO & 232 & 1.3 & 0.294 & 68 & 256 & 134 \\
LSMO & 345 & 1.4 & 0.139 & 54 & 41 & 36\\
\end{tabular}
\caption{Curie temperatures, charge carrier densities, residual resistivities and fitting parameters of
equation \ref{Kubo} for \lcmo\ and \lsmo\ thin films.}
\label{daten}
\end{table}


\begin{references}

\bibitem{DeTeresa97}J.M. De Teresa, M.R. Ibarra, P.A. Algarabel, C. Ritter, C. Marquina, J. Blasco,
J. Garcia, A. del Moral, and Z. Arnold, Nature\ {\bf 386}, 256 (1997).

\bibitem{Jakob98_2}G.~Jakob, W.~Westerburg, F.~Martin, and H.~Adrian,
Phys.\ Rev.\ B\ {\bf 58}, 14966 (1998).

\bibitem{Alexandrov99}A.S.~Alexandrov and A.M.~Bratkovsky, Phys.\ Rev.\ Lett.\ {\bf 82}, 141 (1999).

\bibitem{Jakob98_3}G.~Jakob, F.~Martin, W.~Westerburg, and H.~Adrian,
J.\ Magn.\ Magn.\ Mater\ {\bf 177-181}, 1247 (1998).

\bibitem{Jakob98}G.~Jakob, F.~Martin, W.~Westerburg, and H.~Adrian,
Phys.\ Rev.\ B\ {\bf 57}, 10252 (1998).

\bibitem{Kubo72}K.~Kubo and N.~Ohata, J.\ Phys.\ Soc.\ Jpn.\ {\bf 33}, 21 (1972).

\bibitem{Emin69}D.~Emin and T.~Holstein, Ann.\ Phys.\ (NY) {\bf 53}, 439 (1969).

\bibitem{Snyder96}G.J.~Snyder, R.~Hiskes, S.~DiCarolis, M.R.~Beasley, and T.H.~Geballe,
Phys.\ Rev.\ B\ {\bf 53}, 14434 (1996).

\bibitem{Mandal98}P.~Mandal, K.~B\"arner, L.~Haupt, A.~Poddar, R.~von Helmolt, A.G.M.~Jansen, and P.~Wyder,
Phys.\ Rev.\ B\ {\bf 57}, 10256 (1998).

\bibitem{Khazeni96}
K.~Khazeni, Y.X.~Jia, L.~Lu, V.H.~Crespi, M.L.~Cohen, and A.~Zettl,
Phys.\ Rev.\ Lett.\ {\bf 76}, 295 (1996).

\bibitem{Tokura94}
Y.~Tokura, A.~Urushibara, Y.~Moritomo, T.~Arima, A.~Asamitsu, G.~Kido, and N.~Furukawa,
J.\ Phys.\ Soc.\ Jpn.\ {\bf 63}, 3931 (1994).

\bibitem{Karplus54}
R.~Karplus and J.M.~Luttinger, Phys.\ Rev.\ {\bf 95}, 1154 (1954).

\bibitem{Ye99}
J.~Ye, Y.B.~Kim, A.J.~Millis, B.I.~Shraiman, P.~Majumdar, and Z.~Te\u{s}anovi\'c,
cond-mat/9905007 Preprint (1999).

\bibitem{Friedman63}
L.~Friedman and T.~Holstein, Ann. Phys. (N.Y.) {\bf 21}, 494 (1963).

\bibitem{Jaime97}M.~Jaime, H.T.~Hardner, M.B.~Salamon, M.~Rubenstein, P.~Dorsey, and D.~Emin,
Phys.\ Rev.\ Lett.\ {\bf 78}, 951 (1997),
J. Appl. Phys. {\bf 81}, 4958 (1997).

\bibitem{Asamitsu98}A.~Asamitsu and Y.~Tokura, Phys.\ Rev.\ B {\bf 58}, 47 (1998).

\bibitem{Matl98}P.~Matl, N.P.~Ong, Y.F.~Yan, Y.Q.~Li, D.~Studebaker, T.~Baum, and G.~Doubinina,
Phys.\ Rev.\ B {\bf 57}, 10248 (1998).

\bibitem{Pickett97}
W.E. Pickett and D. J. Singh, Phys.\ Rev.\ B\ {\bf 55}, R8642 (1997).

\bibitem{Wagner98}
P. Wagner, I. Gordon, A. Vantomme, D. Dierickx, M.J. van Bael, V.V. Moshchalkov, and Y. Bruynseraede,
Europhys.\ Lett.\ {\bf 41}, 49 (1998).

\bibitem{Ziese99}M.~Ziese and C.~Srinitiwarawong, Europhys.\ Lett.\ {\bf 45}, 256 (1999).

\bibitem{Chun99}
S.H. Chun, M.B. Salamon, and P.D. Han, Phys.\ Rev.\ B {\bf 59}, 11155 (1999).

\bibitem{Radaelli95}
P.G. Radaelli, D.E. Cox, M. Marezio, S.-W. Cheong, P.E. Schiffer, and A.P. Ramirez,
Phys.\ Rev.\ Lett.\ {\bf 75}, 4488 (1995).

\end{references}
\end{document}